\documentclass[11pt]{article}
\usepackage{natbib,amssymb}
\usepackage{graphicx}
\begin{document}

     \title{\begin{Large}\textbf{Performance of the Micromegas detector in the CAST experiment}\end{Large}}
     \author{S. Aune$^a$, T. Dafni$^b$, G. Fanourakis$^c$, E. Ferrer Ribas$^a$,\\
     T. Geralis$^c$, A. Giganon$^a$, Y. Giomataris$^a$, I.G. Irastorza$^a$,\\
     K.S. Kousouris$^c$, K. Zachariadou$^c$}
     \date{}
     \maketitle
     \begin{center}
     \noindent $^a$DAPNIA, Centre d'etudes de Saclay, Gif sur Yvette Cedex 91191, France\\
     $^b$Institut f\"ur Kernphysik, Technische Universit\"at Darmstadt,  Schlossgartenstr. 9,  64289 Germany\\
     $^c$NCSR Demokritos, Agia Paraskevi 15310, Athens, Greece
     \end{center}
     \begin{abstract}
       The gaseous Micromegas detector designed for the CERN Axion search experiment CAST, operated smoothly during Phase-I, which included the 2003 and 2004 running periods. It exhibited linear response in the energy range of interest (1-10keV), good spatial sensitivity and energy resolution (15-19\% FWHM at 5.9 keV) as well as remarkable stability. The detector's upgrade for the 2004 run, supported by the development of advanced offline analysis tools, improved the background rejection capability, leading to an average rate $5\cdot 10^{-5}$ counts/sec/cm2/keV with 94\% cut efficiency. Also, the origin of the detected background was studied with a Monte-Carlo simulation, using the GEANT4 package.
     \end{abstract}

   \section{Introduction}
   \indent Axions are hypothetical, neutral, pseudo-scalar particles\cite{pdg} that arise in the context of the Peccei-Quinn solution to the strong CP problem. Like neutral pions they have a two-photon interaction and according to the Primakoff mechanism\cite{primakoff} they can transform into photons in external electric or magnetic fields. Therefore, the Sun would be an axion source through the transformation of thermal photons in the solar core. Actually, the expected axion flux on Earth and their spectrum has been calculated\cite{spectrum1,spectrum2} and is confined in the 1-10 keV interval. The detection of these particles is possible in laboratory magnetic fields\cite{sikivie} by the reverse Primakoff effect and their back-conversion into X-rays.
   \section{CAST}
   \indent The CAST experiment (CERN Axion Solar Telescope) is designed to detect solar axions or to improve the existing limits on their coupling to photons. The axion helioscope constructed at CERN uses a decommissioned, superconducting LHC dipole magnet\cite{cast1} which is 10m long and produces a magnetic field of 9T inside two parallel pipes. The magnet is mounted on a rotating platform and its vertical movement allows the accurate observation of the sun for about 3h during sunrise and sunset while the rest of the day is devoted to taking background data. A possible excess signal in all three different X-ray detectors employed in CAST\cite{ccd, mm3} would be attributed to the axion conversion process which is the only X-ray source through the magnet.\\
   \indent The operation of CAST has been divided in two phases in order to scan the axion phase space (coupling to photons vs axion mass): during Phase I the magnet pipes were kept in vaccum while in Phase II they will be filled with buffer gas (He) to look for heavier axions.
   \section{The micromegas detector}
\begin{figure}[ht]
   \centering
   \includegraphics[scale=0.6]{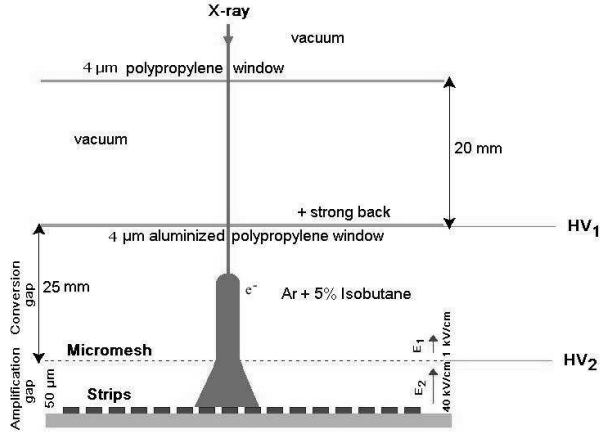}
   \caption{\footnotesize Micromegas detection principle, model V3}
   \label{mm-principle}
\end{figure}

   \indent The micromegas technology (MICRO-MEsh-GAseous Structure) was developed in mid 90's in Saclay\cite{mm1} and is based on a two stage parallel plate avalanche chamber. A micromesh separates the conversion space (2-3 cm), where the primary interaction takes place, from the amplification gap (50-100 $\mu$m) where charge multiplication up to $10^4$ is easily achieved. Such a gain is made possible due to the high electric field applied in the amplification gap, while the large field ratio allows for 100\% electron transmission through the mesh as well as the fast collection of positive ions (100-200 nsec). Moreover, the mesh plane is made of Copper and its allignment is achieved by Kapton pillars spaced 1 mm apart.\\
   \indent The charge collection plane consists of 192 X and 192 Y strips, formed by interconnecting pads on Kapton foil with 350 $\mu$m pitch and their readout is based on Gassiplex chips\cite{mm2}. Due to this two dimensional structure, excellent spatial sensitivity is achieved which can be further improved, according to the experimental needs, by appropriate choice of the strips' size, the gas mixture and the conversion gap\cite{mm4}.\\
   \indent Following the particular demands of the CAST experiment, the micromegas models used were filled with Argon/Isobutane mixture (95\%-5\%) in atmospheric pressure and were supplied with an aluminized polypropylene window (Fig.\ref{mm-principle}), supported by a strong-back\footnote{Made of stainless steel to withstand the vaccum in the magnet bore.}. Also, the frame of the detector was made of low natural radioactivity materials (plexiglas and plastic) to reduce the background,  while the chain of electronics was extended by the installation of a high sampling VME Digitizing Board (MATACQ) to  record the time structure of the mesh pulses. It has 12 bit capacity and is able to handle up to 300 MHz input signals with 2GHz sampling frequency, while producing low noise (less than 0.2 mV rms).\\
   \section{Performance during Phase I}
   \indent CAST Phase I has been succesfully completed and included two running periods during 2003 and 2004, roughly from May to November. The 2003 data have been evaluated and the first results have been published\cite{cast2}, while the 2004 data analysis is still in progress.
     \subsection{2003 run}
\begin{figure}[ht]
   \centering
   \includegraphics[scale=0.4]{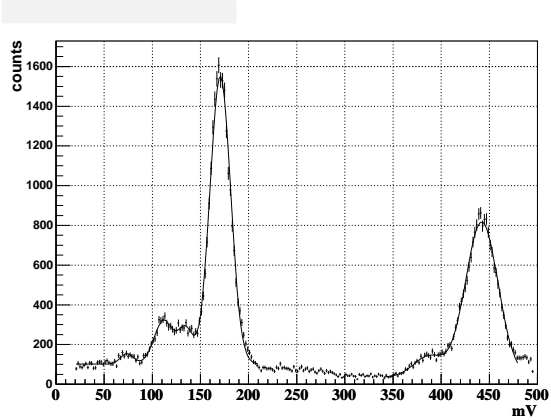}
   \caption{\footnotesize Micromegas responce to Cd source. The peaks were used with ordinary calibration data for the linearity check.}
   \label{Cd_spectrum}
\end{figure}

\begin{figure}[ht]
   \centering
   \includegraphics[scale=0.4]{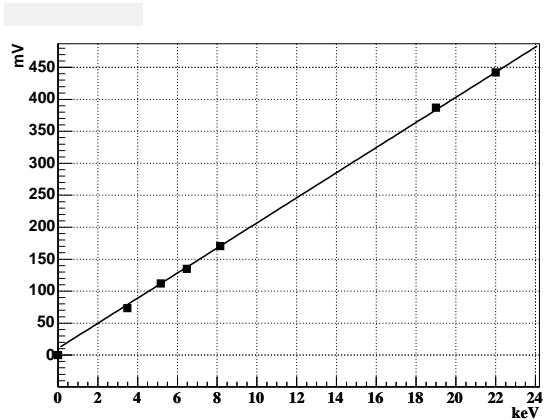}
   \caption{\footnotesize Pulse height vs energy}
   \label{linearity}
\end{figure}

     \indent The V3 micromegas model was designed for the 2003 run of CAST, with 25 mm conversion space and 50 $\mu$m amplification gap. However it was soon realized that the strip structure had some faults, containing a few damaged strips and giving rise to "cross-talk" effects. In oreder to overcome the problem and extract time information of the events, the MATACQ card was installed and which permitted the succesfull operatiom of the detector for the last thee months of the run. Everyday routine included calibration measurement with $^{55}Fe$ source, tracking of the sun and accumulation of background data\footnote{Background is defined as real X-ray events on the detector while it is not tracking the sun.}. The offline analysis was based on sequential cuts on pulses' observables (risetime, width, height-integral correlation), utilizing the calibration data and it has been possible to reduce the background rate\footnote{The net rate was approximately 1 Hz.} to the average value of $1.4\cdot10^{-4} sec^{-1}cm^{-2}keV^{-1}$ in the 1-9 keV region with 80\% and 95\% cut efficiency at 3 keV and 5.9 keV respectively.\\
     \indent Moreover, the detector's linear responce was verified (Fig.\ref{linearity}) by using $^{109}Cd$ source which produced fluorescence of the device's materials (Fig.\ref{Cd_spectrum}). The energy resolution was 15\% FWHM at 5.9 keV and the system's stability has been outstanding since the time characteristics of the mesh pulses as well as the energy response variated less than 2\% through the whole running period.\\
\\
     \subsection{2004 run}
\begin{figure}[ht]
   \centering
   \includegraphics[scale=0.4]{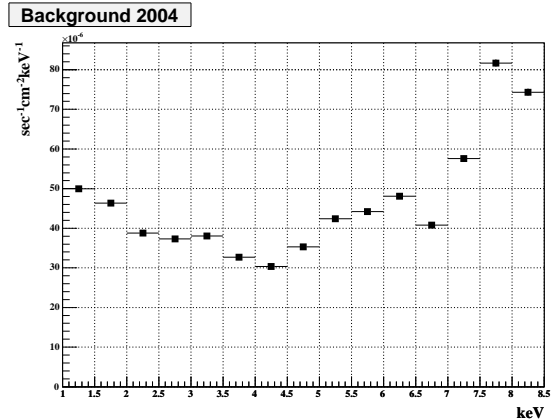}
   \caption{\footnotesize 2004 background shape}
   \label{Bkg_2004}
\end{figure}

     \indent The experience acquired during the 2003 run led to the development of the V4 model (30 mm conversion gap, 100 $\mu$m amplification gap) which was specially designed to eliminate the "cross-talk" effects present at the previous model and to improve the quality of the strips. As a result, both goals were achieved\footnote{The damaged strips were less than 1.3\%.} and a faster MATACQ board was installed, reducing the detector's dead time to 14 msec (less than 1.5\% of the net data rate) while the energy resolution was 19\% FWHM at 5.9 keV.\\
     \indent The offline analysis was greatly improved by combining the information from the spatial distribution of the charge produced after an event with the time structure of the mesh pulses. More specifically, six observables (risetime, width, height vs integral correlation, X and Y strip multiplicity balance, X and Y strip charge balance, height vs total strip charge correlation) were used in a Fisher discriminant method to distinguish more efficiently the proper X-ray events from other signals. The resulting background rejection was $4.8\cdot10^{-5}sec^{-1}cm^{-2}keV^{-1}$ in the 1-8.5 keV region (Fig.\ref{Bkg_2004}) with 94\% uniform software efficiency.\\
     \indent The system's stability is demonstrated through the mesh pulses' time structure (0.5\% variation of risetime and width) and the moderate gain variation (10\% on a week scale) which was corrected with everyday calibration.\\
\\
\indent In order to understand the origin of the measured background, the GEANT4 package was used for a Monte-Carlo simulation. As a first approach, the simulation included the responce of the detector to incident particles without the full reconstruction of an event which can be achieved by more specialised simulation toolkits. The first results indicate that the background is induced by environmental radioactivity\footnote{Dominated by $^{222}Rn$.} and gamma radiation which produce fluorescence of the surrounding materials \footnote{Primarily Cu from the detector's mesh and also Fe from the window's strongback}. It was also revealed that the thermal neutrons present in the experimental site interact via elastic scattering with the Ar atoms of the detector's gas, the recoil of which gives identical signal to an X-ray event. Other ionizing particles, such as muons or electons are also detected but easily identified due to their non-local energy deposition. The quantitative reproduction of the background has not been possible because of limited knowledge about the rate and the spectrum of the surrounding radiation but a detailed measurement has been scheduled, which will be used in the CAST Phase II.
   \section{Conclusions and prospects}
\indent The micromegas technology has been employed for the construction of a reliable
X-ray detector matching the demanding requirements of CAST. The expectation for low background and adequate stability throughout the extended running periods of the experiment was completely met by micromegas which exhibited at the same time good energy resolution.\\
\indent Currently, a new micromegas detector is being developed in order to operate during CAST Phase II with an X-ray focusing device. On the other hand, Monte Carlo studies are under way to investigate the optimum shielding needed for further reduction of the background.
   \section{Acknowledgements}
We would like to thank all the members of the CAST collaboration for their work during the Phase-I of the experiment.
   \bibliographystyle{plain}
   
\end{document}